\documentclass{kapproc} 
\usepackage[dvips]{graphicx}
\upperandlowercase
\kluwerbib 
\begin{document}

\def\ga{\mathrel{\hbox{\rlap{\hbox{\lower4pt\hbox{$\sim$}}}\hbox{$>$}}}}
\def\la{\mathrel{\hbox{\rlap{\hbox{\lower4pt\hbox{$\sim$}}}\hbox{$<$}}}}

\articletitle[AM CVn Stars in the UCT CCD CV Survey.]
{AM CV{\small n} Stars in the UCT CCD CV Survey}

\author{Patrick A.~Woudt and Brian Warner}

\altaffiltext{}{Department of Astronomy,
University of Cape Town, Rondebosch 7700, South Africa}
\email{pwoudt@circinus.ast.uct.ac.za, warner@physci.uct.ac.za}

\begin{abstract}
High speed photometry of the helium-transferring binary ES Cet -- 
taken over a two-year period (2001 October -- 2003 October) -- shows a very stable 
photometric period of 620.211437 $\pm$ 0.000038 s, with a tentative indication of curvature in the 
O--C diagram suggesting a change in period at a rate of $\dot{P} \sim 1.6 \times 10^{-11}$.
Phase-resolved spectroscopy of ES Cet obtained
with the Hobby-Eberly Telescope shows a clear modulation on the photometric period, the assumed
orbital period. We have followed a newly identified AM CVn star (`2003aw') 
photometricaly through its 2003 February/March outburst during which it varied
in brightness over a range of V = 16.5 -- 20.3; we find a superhump period of 2041.5 $\pm$ 0.3~s. Questions are raised
about the reality of the detected spin-up in RX\,J0806 (Hakala et al.~2003; Strohmayer 2003).

\end{abstract}


\section{Introduction}

There are currently ten known unequivocal double degenerate interacting binaries (AM CVn stars), namely
ES Cet, AM CVn, HP Lib, CR Boo, KL Dra, V803 Cen, CP Eri, `2003aw', GP Com and CE-315,
ranging in orbital period ($P_{orb}$) from 10.3 -- 65.1 min. These stars have proper spectroscopic
and photometric credentials -- their spectra show helium emission or absorption lines; no hydrogen can be present in these systems. 
There are two additional candidate AM CVn stars of suspected short orbital period, RX\,J0806 at $P_{orb}$ = 5.35 min
(Israel et al.~2002; Ramsay et al.~2002) and V407 Vul (Cropper et al.~1998) at $P_{orb}$ = 9.49 min. 
Their classification as AM CVn stars is, however, not unambigious; there is some (tentative) evidence for the presence of hydrogen in the
spectrum of RX\,J0806 (Israel et al.~2002), and the spectrum of V407 Vul is that of a K star (Steeghs 2003)
making it appear like an intermediate polar precursor at quite long orbital period (Warner 2003). In this interpretation,
the 9.49-min photometric and X-ray modulation is associated with the spin period of the primary, not the orbital period.
Table 1 lists all the AM CVn stars, including the two candidates.

\begin{table}[ht]
\caption[The AM CVn stars]
{The AM CVn stars}
\begin{tabular*}{\textwidth}{@{\extracolsep{\fill}}lllll}
\sphline
\it Object&\it V  (mag)&\it $P_{orb}$ (s)& \it $P_{sh}$ (s)& \it References \cr
\sphline
RX\,J0806 & 21.1 & 321.25$^a$  &        & 1, 2\cr
V407 Vul  & 19.9 & 569.38$^a$  &        & 3\cr
ES Cet    & 16.9 & 620.21144     &        & 4, these proceedings \cr
AM CVn    & 14.1 & 1028.7      & 1051.2 & 5, 6\cr
HP Lib    & 13.7 & 1102.7      & 1119.0 & 7, 8\cr
CR Boo    & 13.0 -- 18.0 & 1471.3 & 1487 & 9, 10\cr
KL Dra    & 16.8 -- 20   & 1500 & 1530  & 11 \cr
V803 Cen  & 13.2 -- 17.4 & 1612.0 & 1618.3 & 12 \cr
CP Eri    & 16.5 -- 19.7 & 1701.2 & 1715.9 & 13 \cr
`2003aw'  & 16.5 -- 20.3 &        & 2041.5 & 14, these proceedings \cr
GP Com    & 15.7 -- 16.0 & 2974   &        & 15, 16 \cr
CE-315    & 17.6         & 3906   &        & 17, 18 \cr
\sphline
\end{tabular*}
\begin{tablenotes}
$^a$Not yet definitively established as orbital periods.\newline
$^1$Israel et al.~(2002); $^2$Ramsay et al.~(2002); $^3$Cropper et al.~(1998); $^4$Warner \& Woudt (2002);
$^5$Solheim et al.~(1998); $^6$ Skillman et al.~(1999); $^7$O'Donoghue et al.~(1994); $^8$Patterson et al.~(2002);
$^9$Wood et al.~(1987); $^{10}$Patterson et al.~(1997); $^{11}$Wood et al.~(2002); $^{12}$Patterson et al.~(2000);
$^{13}$Abbott et al.~(1992); $^{14}$Woudt \& Warner (2003a); $^{15}$Nather et al.~(1981); $^{16}$Marsh et al.~(1991);
$^{17}$Ruiz et al.~(2001); $^{18}$Woudt \& Warner (2002).
\end{tablenotes}
\end{table}
\inxx{captions,table}


\section{The UCT CCD CV Survey}

The UCT CCD CV Survey is a high speed photometric survey of faint cataclysmic variable stars (CVs) 
using the University of Cape Town (UCT) CCD photometer (O'Donoghue 1995)
in frame-transfer mode, in combination with the 1.0-m and 1.9-m 
reflectors at the Sutherland site of the South African Astronomical Observatory.

\subsubsection{ES Cet}

Initial high-speed photometry of ES Cet obtained during four nights in 2001 October (Warner \& Woudt 2002)
showed a clear modulation at 620.26 s -- in the Fourier transform only the fundamental and its first three
harmonics of the 620.26-s modulation were present. The spectrum of ES Cet (see Fig.~\ref{spectESCet}) is dominated
by He\,II emission lines, and hence its position amongst the AM CVn stars is secure. From the low mass ratio
($q$ = 0.094), and the predicted rate of mass transfer $\dot{M}$ 
of $\sim 1 \times 10^{-8}$ M$_{\odot}$ y$^{-1}$ at $P_{orb}$ = 620 s (Warner 1995), 
one expects that the photometric modulation originates
from superhumps due to tidal distortions in the accretion disc (Patterson et al.~2002). In this case, the orbital
period would be a few per cent lower than the observed photometric modulation.

\begin{figure}[ht]
\centerline{\includegraphics[width=10.5cm]{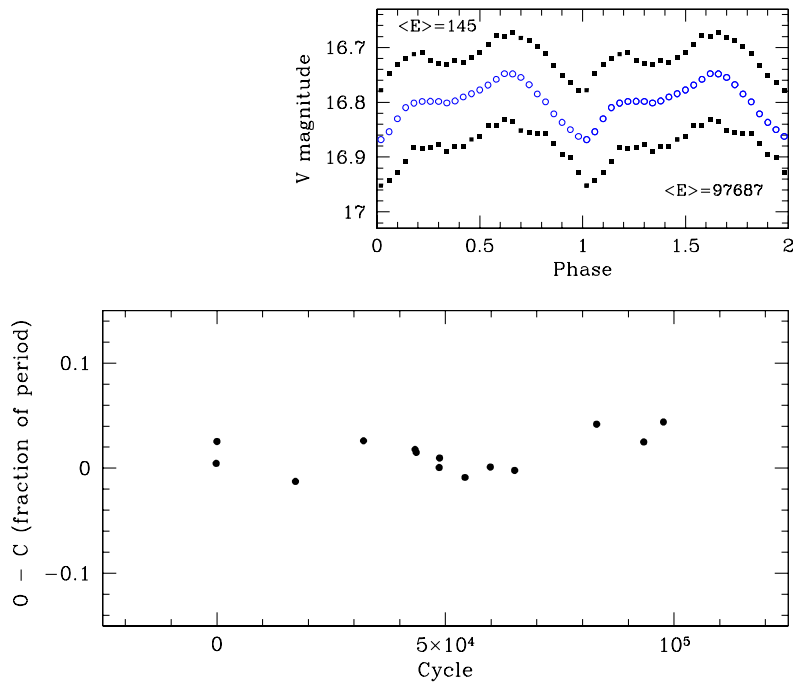}}
\caption{The O--C diagram (lower panel) of all the UCT CCD photometry obtained of ES Cet over the last 2 years. The upper panel
shows the average light curve of all the data (open circles) compared with the mean light curve of two individual runs spaced two
years apart.}
\label{escetoc}
\end{figure}

\begin{figure}[ht]
\centerline{\includegraphics[width=11cm]{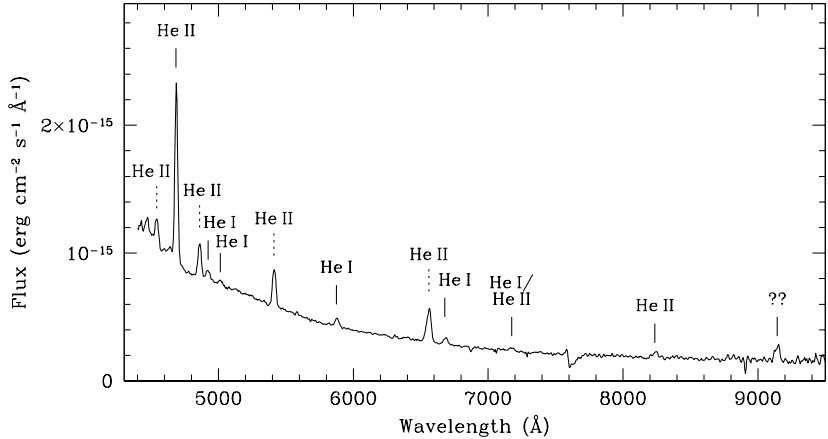}}
\caption{The averaged spectra of ES Cet taken with the Hobby-Eberly telescope.}
\label{spectESCet}
\end{figure}

We have followed ES Cet photometrically over the last two years and the photometric modulation
is surprisingly stable. If indeed it arises from a superhump modulation, it is the most stable superhump
detected to date (Patterson, priv. comm.). The O--C diagram of all the UCT CCD
photometry taken between 2001 October and 2003 October is shown in Fig.~\ref{escetoc}, phased according to
the ephemeris given in Eq.~\ref{ephescet}. Even though there
is some scatter in the O--C diagram, there are no substantial phase shifts or period changes (as might have been expected
were the modulation due to superhumps). 

\begin{equation}
{\rm HJD_{min}} = 245\,2203.3739512 + 0.0071783731 (4) \, {\rm E}
\label{ephescet}
\end{equation}

With the two-year baseline, there is a slight hint of upwards curvature in the O--C diagram, which 
implies a lengthening of the photometric period.  If the data are split in two halves (E $\sim$ 0 -- 50\,000 cycles and 
E $\sim$ 50\,000 -- 100\,000 cycles, with an overlap of the dense data coverage at E $\sim$ 50\,000 cycles), we find that the
period is indeed larger in the second half: $P_{orb}$ (1st year) = 620.211391 ($\pm$ 50) s versus $P_{orb}$ (2nd year) = 620.211841 
($\pm$ 96) s. The amount of variation is consistent with the curvature seen in the lower panel of Fig.~\ref{escetoc}.
It implies $\dot{P} \sim 1.6 \times 10^{-11}$ -- fairly close to the expected rate of change of $6 \times 10^{-12}$ (Warner \& 
Woudt 2002) for a high $\dot{M}$ system. 
We realise that the data coverage is still rather small, and another two years of photometry will be required
to confirm this trend in the O--C diagram. 

\begin{figure}[ht]
\centerline{\includegraphics[width=11cm]{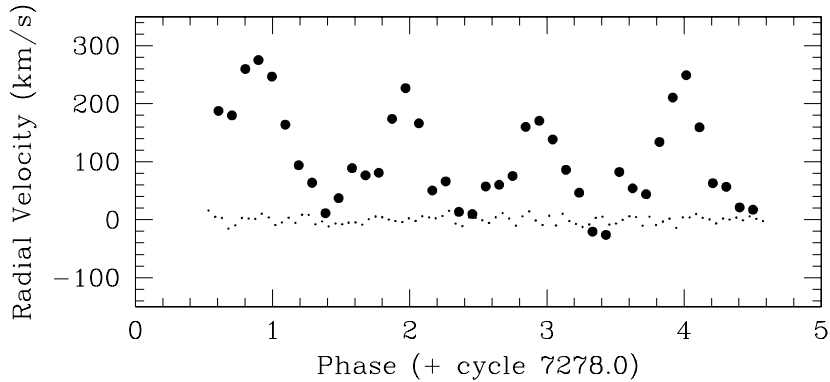}}
\caption{Variations in the centroid of the He\,II 4686 {\AA} emission line (big dots), phased on the ephemeris given in Eq.~\ref{ephescet}.
The small dots show the [O\,I] 5577 {\AA} night sky line.}
\label{spectvar}
\end{figure}

Apart from the extended photometric coverage, we have obtained phase-resolved spectroscopy (with a time resolution of 30 s) 
of ES Cet using the Low Resolution Spectrograph on the Hobby-Eberly Telescope (HET) at the McDonald Observatory in Texas. Two visits of
40 minutes, and a third observing run 30 minutes long, showed very clearly that the spectral lines varied on the photometric
period of 620.21 s. This has also been seen by Steeghs (2003) in two consecutive nights of phase-resolved Magellan 
data. The averaged HET spectra (combining all the spectra of the three different ES Cet observations) is shown in 
Fig.~\ref{spectESCet}.
Fig.~\ref{spectvar} shows the variation of the centroid of the He\,II 4686 {\AA} emission line as a function of the photometric
ephemeris given in Eq.~\ref{ephescet}. 

The spectrum consists mainly of He\,I and He\,II emission lines (lines of the He\,II Pickering series are marked by the dashed vertical
bars in Fig.~\ref{spectESCet}); there is one line at $\lambda \sim 9140$ {\AA} which we haven't
yet been able to identify. The strong emission lines are somewhat unexpected for an object of inferred high $\dot{M}$.
The low spectral resolution of the HET spectra fails to show the double-lined nature of the emission lines, but higher resolution
spectra (Steeghs 2003) clearly show the double emission lines, indicating the presence of an accretion disc.
The spectral resolution of the HET spectra is too low for generating Doppler tomograms, cf. Steeghs (2003).

Given the stability of the photometric period and spectroscopy modulation on the photometric period, it seems that 
the 620-s modulation is more probably
the orbital period of the system and not the superhump period as commonly expected for a low-mass ratio, high $\dot{M}$
system. 

\subsubsection{`2003aw'}

Shortly after its outburst in 2003 February, `2003aw' was identified as a candidate supernova in a supernova
search (Wood-Vasey et al.~2003), but a spectrum taken by Chornock \& Filippenko (2003) revealed weak He\,I emission lines at zero
redshift, making `2003aw' a candidate AM CVn star caught in a high state. Initial photometry with the UCT CCD confirmed
this suggestion after finding a photometric modulation in the light curve with a period of 2034 $\pm$ 3 s (Woudt \& Warner 2003b),
with recurring dips in the light curve during the high state -- possibly indicating shallow eclipses.

We followed `2003aw' through its decline from the high state into the intermediate state, and during both phases
a photometric modulation of identical period was found: P$_{sh}$ = 2041.5 $\pm$ 0.3 s (Woudt \& Warner 2003a).
The dips did not occur during the intermediate state. However, during the intermediate state we observed a 
`cycling' in brightness of $\ge 0.4$ mag on a time-scale of $\sim$ 16 h. This behaviour is also seen in other AM CVn stars
in intermediate states; in CR Boo ($P_{orb} = 1471$ s) the cycle time is $\sim$ 19 h, with a range of 1.1 mag and in V803 Cen
($P_{orb} = 1612$ s) the cycle time is 22 h, with a range of 1.1 mag. `2003aw' has two other interesting aspects: 

\begin{itemize}
\item{The presence
of sidebands to the fundamental {\sl superhump} frequency and two of its harmonics during the high state. 
The frequency separation of these sidebands corresponds roughly to the `cycling' time scale.}
\item{A short lived ($\sim$ 1 day) brightening of $\Delta$V $\sim$ 1.8 mag occured during the high state, resembling
the behaviour of some intermediate polars (Schwarz et al.~1988; van Amerongen \& van Paradijs 1989).}
\end{itemize}

`2003aw' seems to know its place within the emerging hierarchy of AM CVn stars. The systems of shortest orbital
periods ($P_{orb} \la 1200$ s) have stable high $\dot{M}$ discs, systems with periods between $\sim$ 1200 -- 2500 s -- to which 
`2003aw' belongs -- have unstable high $\dot{M}$ discs (the equivalent of the nova-likes of VY Scl type), and systems with
orbital periods $P_{orb} \ga 2500$ s have low $\dot{M}$, and are perhaps permanently in a low state.

\section{On the spin-up in RX\,J0806}

Two recent papers (Hakala et al.~(2003); Strohmayer (2003)) presented evidence that the
orbital period in RX\,J0806 is undergoing a spin up. Evidence for this was based on three epochs of 
data: X-ray data taken with ROSAT in 1994-1995 (Burwitz \& Reinsch 2001), and two sets of optical 
data taken with the VLT and NOT (Hakala et al.~2003) in 2001 Nov/2002 Jan, and 2003 Jan/Feb, respectively. 
Of the two optical data sets, the first data set (2001/2002) suffers from severe aliasing (Hakala et al.~2003), and
as a result, the deduced period evolution depends critically on the assumption that the X-ray period in the 1994/1995
ROSAT data set is correct.

\begin{figure}[ht]
\centerline{\includegraphics[width=11cm]{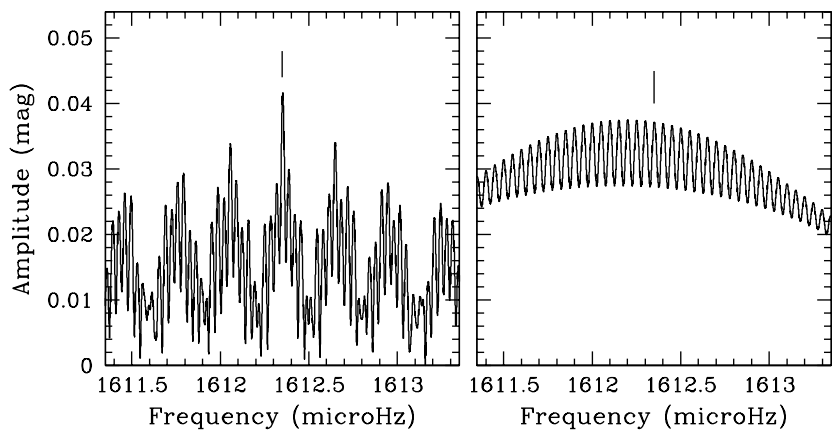}}
\caption{A comparison of the Fourier transform of all the SAAO observations of ES Cet in the period 2001 October --
2003 October (left panel) with a selected subsample with limited data coverage the equivalent of 80 cycles over 32\,000
cycles (right panel). The proper photometric period is marked by the vertical bar in both diagrams.}
\label{testft}
\end{figure}

The X-ray data (Burwitz \& Reinsch 2001) were taken in 1994 October and 1995 April with a total of 13\,400 s of
integration time. This amounts to the equivalent of 42 cycles of the 5-min modulation spread out over $\sim$ 180 days
($\sim$ 50\,000 cycles). With such poor data coverage, it is impossible to determine periods to the accuracy of
0.4 ms as quoted in Burwitz \& Reinsch (2001) and perpetuated in Hakala et al.~(2003). The aliasing is severe,
as shown in Figure 4 of Burwitz \& Reinsch (2001), and the highest peak in the forest of aliases is not necessarily 
the correct period. Each alias peak can be determined with a precision of 0.4 ms (largely determined by the baseline of the observations), but
the choice of alias peak can lead to inaccurate results.
To illustrate that, we have selected a few observing runs of ES Cet -- mimicking an approximately 
similar data coverage (80 out of 32\,000 cycles) -- and show the FT for this small data set next to the complete
data set in Fig.~\ref{testft}.

If the nearest alias to the right (see Figure 4 of Burwitz \& Reinsch) of the preferred period in Burwitz \& Reinsch (2001) 
is chosen (i.e., a lower period), the entire period evolution disappears (Figure 2 of Hakala et al.~2003). At best, the 
evidence for a spin-up in RX\,J0806 is tentative; there are currently insufficient data to claim (with high 
significance) a period evolution in this object. It remains to be seen if the proposed period evolution in V407 Vul (Strohmayer 2002)
suffers from the same problem of poor data coverage (in that case there is X-ray data for 172 cycles spread over 250\,000 cycles
of the 569-s modulation).

\section{Discussion}

To determine the period evolution of short period AM CVn stars a dedicated long-term campaign is needed in order 
to eliminate aliases and cycle count uncertainties. After two years of observations of ES Cet, the O--C diagram is starting to show
a slight upwards curvature (indicating a increase in period)
and we derive a (tentative) value for $\dot{P} \sim 1.6 \times 10^{-11}$, or the equivalent
$\dot{\nu} \sim -4 \times 10^{-17}$ Hz s$^{-1}$. Insecure though it is, it may be the first detection of a
$\dot{P}$ in an AM CVn system.

\begin{acknowledgments}
PAW's research if funded by a strategic grant from the University of Cape Town
and by funds from the National Research Foundation. BW's research is funded by the 
University of Cape Town. We kindly thank the HET Board for granting observing time.
\end{acknowledgments}

\end{document}